# Analyzing 50 years of major fog events across the central coastal plain of Israel

Noam David[1], Asaf Rayitsfeld[2], H. Oliver Gao[3*]

1. AtmosCell, Tel Aviv, Israel (www.atmoscell.com)
2. The Israeli Meteorological Service, Beit-Dagan, Israel
3. The School of Civil and Environmental Engineering, Cornell University, Ithaca, NY,

*Correspondence to:* H. Oliver Gao, e mail: hg55@cornell.edu

**Abstract**

This report presents an analysis of 152 major fog events that have been occurring for five decades (1967-2017) across the central coastal plain of Israel. Analysis of the meteorological data shows that fog events in the experimental area predominantly occur under two sets of synoptic conditions – Red Sea Trough (44%) and Ridge (41%), while the incidence of fog events peaks between March and June. In particular, the results obtained indicate a decreasing trend in the number of fog events and their duration over time where the frequency of radiation fog has decreased over time when compared to the incidence of advection fog. This note provides a long-term analysis of data in a region that lacks reliable time series of this length, and highlight important insights for future research.

**Introduction**

According to the American Meteorological Society, fog is defined as a state where water droplets suspended in the air near the Earth's surface reduce visibility below 1 km (AMS, 2020). The economic damages caused due to fog can be vast, and on a scale comparable to the damages from winter storms, as a result of the disruption to flight schedules, for example, or even the shutting down of airports in severe cases (Gultepe et al., 2009). By way of demonstration, research has found that accurate timing of the onset and ending of capacity limiting situations such as fog could save busy air terminals around New York City approximately $480,000 per event (Allan et al., 2001).

Additional negative effects related to the phenomenon include acid fog that can cause damage to vegetation and structures, and smog, which presents a health risk, particularly for people suffering from respiratory illnesses (e.g. Tanaka et al., 1996; Wichmann et al., 1989).

However, the phenomenon also has positive contributions. Thus, for instance, in areas with low water availability, fresh water for afforestation, gardening, and even potable water can be harvested from fog (Klemm et al., 2012). Additionally, fog plays an important role in scrubbing the atmosphere through particle scavenging and drop deposition processes (Herckes et al., 2007).

Tools for monitoring fog include a variety of ground level sensors, human observers and satellite systems (David et al., 2013).

Human observers estimate the visibility during fog based on the appearance or obscuring of objects located at predetermined distances from the observers location. The observations derived by this method, though, are not objective, since one observer's visibility estimate might be different from another observer's estimate.

Satellites provide wide spatial coverage and map areas where fog exists (e.g. Lensky, and Rosenfeld, 2008), but, at times, do not provide sufficient response, for example, due to obscuring of the fog from the satellite's viewpoint as a result of high altitude cloud cover. Additionally, satellite systems cannot, at times, differentiate, for example, between a low stratus cloud, at an elevation of several tens of meters above ground level, that does not endanger drivers, and fog that lies in immediate proximity to the ground (e.g. Gultepe et al., 2007; David, 2018).

On the other hand, specialized sensors such as visibility meters, Runway Visual Range (RVR) sensors and particle monitors can provide precise and reliable fog observations near ground level, however these instruments can only provide a local measurement, that does not reliably represent the entire space.

Hence, reliable monitoring of the phenomenon, over a wide geographic area, is still currently a challenge, and efforts are being made to develop tools for mapping fog in high temporal and spatial resolution using alternative and complimentary solutions (e.g. David et al., 2015; 2019).

Categorizing broadly, fog can be classified as either radiation fog or advection fog, based on the physical processes that cause it to develop (Ziv and Yair, 1994). The first is caused as the result of radiative cooling of the surface, and the optimal conditions for its creation including a clear night, allowing for efficient radiative cooling, light wind (less than 5 knots), stability, and humid air. Advection fog is caused by relatively warm air being cooled to saturation as a result of it being carried by a light wind over a cold surface. The optimal conditions for its creation include a 5-10 knot wind and atmospheric stability. At times, a combination of both of these processes can cause the creation of fog, for example in cases where the surface is not cold enough to cause the condensing of droplets in the air traveling over it alone, but where the addition of radiative cooling can lead to the completion of the process.

As has been extensively reviewed by Klemm and Lin (2015), the frequency and intensity of fog events vary greatly over time. Broadly, the majority of research reports, from different locations across the world, indicate a major decrease in the frequency of fog formation, and its intensity. In most of the measuring stations where observations were carried out (e.g. Chen et al., (2006); Vautard et al. (2009); LaDochy and Witiw (2012); Williams et al. (2015)). In some cases an increase was observed (e.g. Syed et al., (2012)).

Trends in fog frequency and intensity can be a result of changes in regional climactic conditions. Urban Heat Island (UHI) effect, or changes in predominant circulation patterns, can lead to increasing air temperatures and a resulting decrease in Relative Humidity (RH), and as long as there are no feedback mechanisms overriding the temperature effect, may alter fog trends. Changes in the number of cloud condensation nuclei (CCN) has been discussed as a potential cause for fog trends, though there has yet to be a discussion of the relevant physical processes behind this reasoning (Klemm and Lin, 2015).

In this study, we present an analysis of fog measurements taken over 5 decades (1967 – 2017) in Israel's central coastal region. Based on analysis of this data, we report a meaningful decrease in the frequency of fog creation and their duration. Additionally, we point out the key synoptic conditions that comprise the mechanism for the creation of fog in the area and analyze some of the characteristics of the phenomenon.

**Classification for advection and radiation fog**

In order to distinguish between the two types of fog we examined the vertical structure of the temperature and the dew point in the lower tropospheric levels up to 850 mb. The classification process was performed on the basis of the radiosonde measurements, launched from Bet Dagan station every night between 23:00 to 00:00 UTC (Figure 1). Depending on the two different profile types, two types of fog were observed.

*Radiation fog* was typically characterized by a deep temperature inversion which extended from the surface level up to a pressure level of about 950 mb (~500 m above sea level). Simultaneously, the dew point was increasing with the increasing temperature, conditions that created a stable moist layer. While under the above conditions winds on the lower two levels were measured to be less than 4 knots (i.e. at ground level and at a height of 1000 millibars), the fog was classified as an radiation fog type. The fog was classified as *advection fog* when measured wind speed was higher than 4 knots (an up to 10 knots) during a low marine inversion and / or a weak ground inversion.

**Results**

We studied 152 fog major events which took place between March 1967 to March 2017 across the test site located in the central coastal plain of Israel. Figure 1 shows the experiment site where meteorological measurements and visibility estimates, acquired by professional human observers, were taken from Beit Dagan surface station. Additional visibility estimates were taken by observers located at Ben Gurion airport.

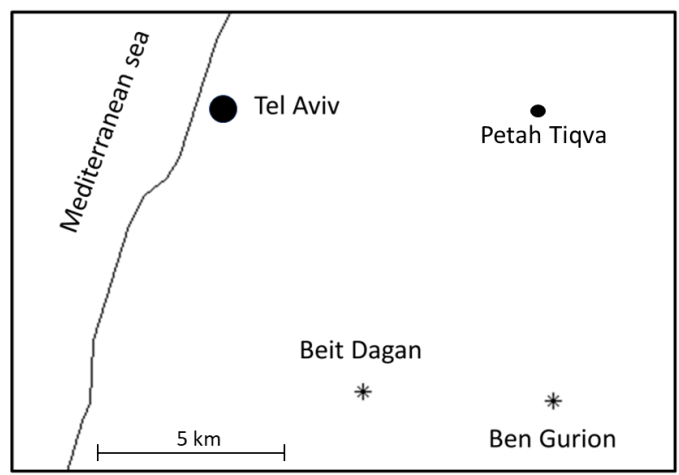

**Figure 1: The experiment site situated in the central costal plain of Israel. The asterisks mark the locations of the surface stations, in the vicinity of the city of Tel Aviv.**

Fog events were determined according to the Israeli Meteorological Service (IMS) database - based on visibility data, relative humidity, wind velocity, radiosonde records and synoptic conditions. The measurements that were available for the entire period were stored in SYNOP code, and accordingly, are available at a sampling frequency of once every three hours. The set of radiosonde measurements analyzed was gathered from nightly releases from the Beit Dagan station. We also note that events documented in the database at only one specific hour were considered as events of 1 hour duration in the calculations. The analysis of the results focused on fog events that we defined as significant, that is, events that were observed by both stations in the same time frame. Thus, an event was defined as a fog event when visibility was estimated by the professional observers at Beit Dagan and Ben Gurion Airport to be less than 1 km simultaneously, or, at times where visibility was estimated to be less than 1 km at one station, under the condition that fog was also detected at the other station within a 6 hour time interval from when it was detected at the first station.

Figure 2 presents general details regarding the fog events that took place across the experiment site between 1967 and 2017, and particularly, the frequency of fog creation given a certain synoptic condition (Fig. 2a), the total number of fog events occurring per month (Fig. 2b) and the distribution of average visibility estimates at each station (Fig. 2c).

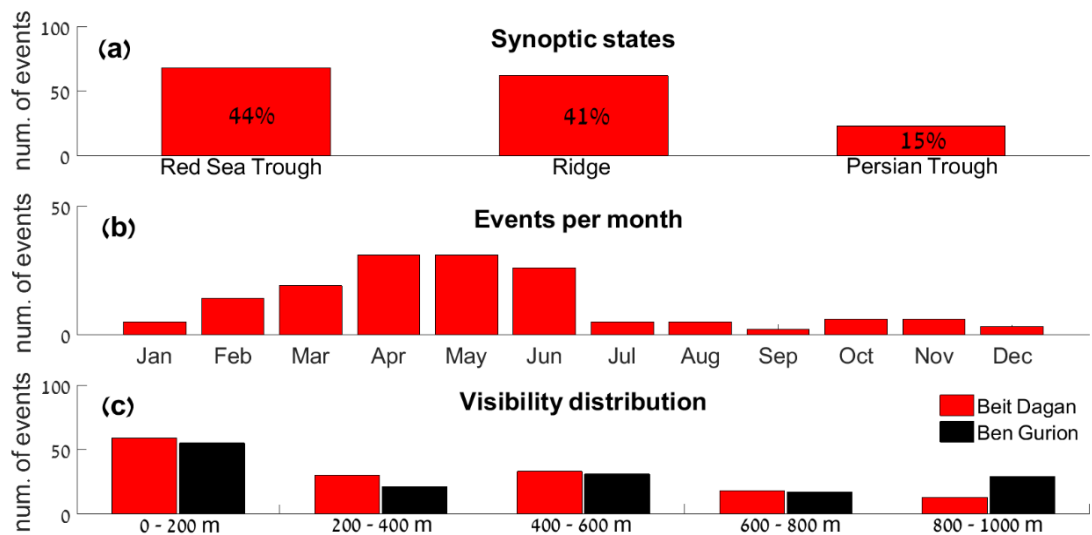

**Figure 2: Analysis of fog events between 1967 and 2017. (a) Frequency during different synoptic conditions. (b) Total major fog events per month. (c) Distribution of average visibility estimates for the Beit Dagan station (red) and Ben Gurion (black). To produce graph (c), the average visibility was calculated for each station separately, from the visibility observations taken during each fog event.**

It can be seen, then, that most of the fog evets in the experiment area occur under conditions of Red Sea Trough (44%) and Ridge (41%), where the incidence of events in the area peaking between March and June. In the major percentage of fog events (observed by both stations, as defined here) average visibility is lower than 600 meters. Figure 3 shows the total number of radiation fog events (Fig. 3a), advection fog (Fig. 3b) and the distribution of the total number of events combined divided into three equal periods.

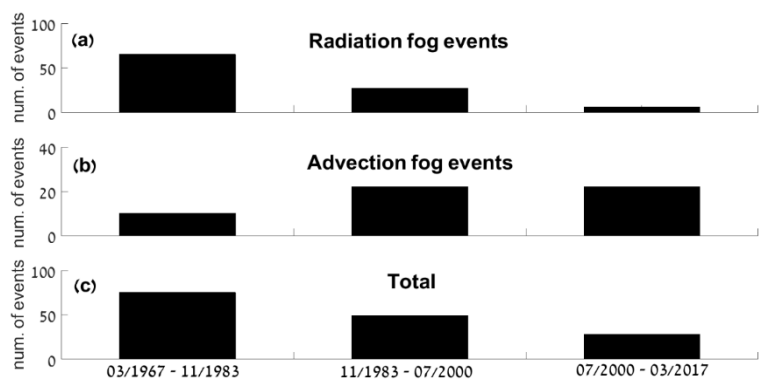

**Figure 3: Radiation and Advection fog. Number of radiation fog events (Fig. 3a), number of advection fog events (Fig. 3b), and total number of fog events (Fig. 3c) divided into 3 equal periods.**

From analysis of the data we found that 64% of fog events analyzed had radiation fog characteristics, while 36% had characteristics of advection fog.

We note, in the overall view, that while the incidence of radiation fog has decreased over time, the incidence of advection fog has increased when compared to the first third of the experiment period. We note that the total number of fog events decreased measurably over time.

We focus, then, on this aspect. Figure 4 shows the trend of fog creation frequency and duration. The linear fit approximations of the fog event records are listed at the top right of each panel. It can be seen that the total number of events per year has generally decreased over time (Fig. 4a) and that the total number of hours where fog existed per year has decreased as well. Fig. 4c, which was constructed from Figures 4b and 4a shows the average duration of a single fog event. The linear fit calculated, indicates a more or less constant (an overall very small decreasing trend) in the average duration of each single event.

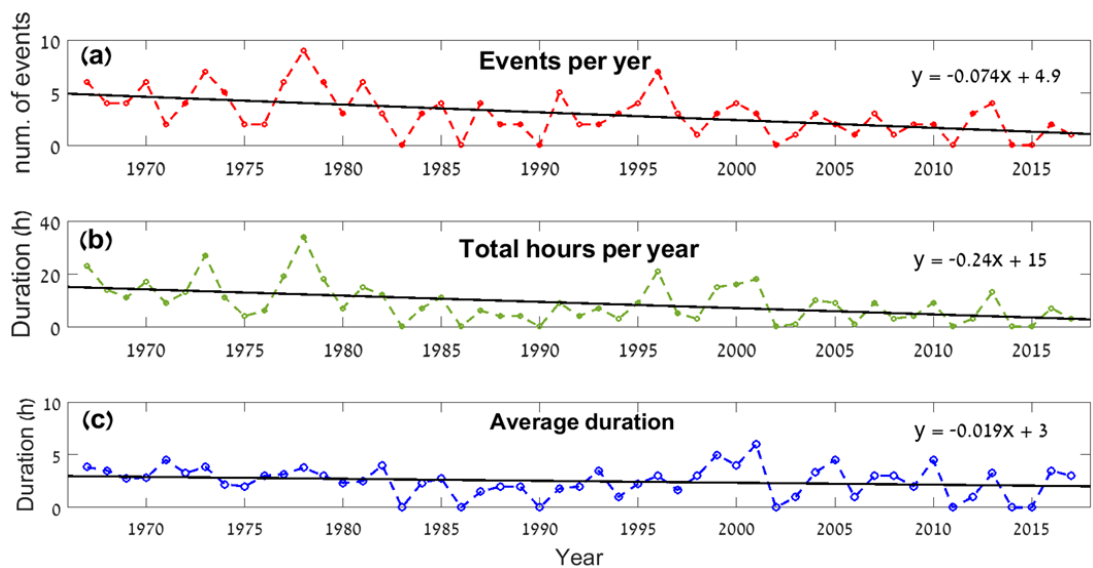

**Figure 4: Incidence and duration of fog per year for the experimental area. Number of meaningful fog events per year (a), total number of fog event hours in every year (b), average duration of fog event per year.**

**Discussion**

It has been shown that a temperature increase of a few tenths of a degree can strongly affect the visual range (Klemm and Lin, 2015). In particular, the temperature increase (and a decrease of the aerosol concentrations) can lead to increased visual range, i.e. a decrease of fog.

Previous climate change observations have already shown temperature increases of a few tenths of a degree over vast parts of the terrestrial surface (Stocker et al., 2013).

Zhang et al. (2005) analyzed data from 75 stations across 15 countries in the Middle East region for the period 1950–2003. Their data analysis showed statistically significant, and spatially coherent, trends in temperature indices pointing to a warming trend in that area. They also reported a significant increase in the frequency of warm days which has been observed towards the 1990s, while, since the 1970s, the frequency of cold days has gradually decreased significantly. More specifically to the experimental area discussed here, recent research conducted on extreme temperature (and precipitation) indices in Israel between 1950 and 2017 – a period that covers the entire period of research of this work – found that temperatures (specifically, the daily minimum / maximum temperatures) are trending upwards (Yosef et al., 2019).

Due to the relationship between the increase in temperature and the increase in visibility range, it is reasonable to assume that the decrease in fog is driven by climate change (Klemm and Lin, 2015). That being said, we note that this is a general statement, and downscaling it to the specific case reviewed here is non-trivial, and requires future research.

Moreover, the observation stations of this study are located in Israel's central coastal plain, a region with temperate weather, at an elevation of between 30-40 meters above sea level. These days, the area where the stations are located is urban, but includes agricultural fields spread within it.

We note that over the years since the experiments beginning and until the writing of this paper, major changes have occurred in the land surface due to intensive construction in the area, and the region has transformed from one with a more rural character, where building, vehicle and population densities were relatively low, to an area where a major transportation intersection is located, one where population and building densities have, accordingly, dramatically increased over time.

Figure 5 (a) shows a photograph of the experiment area from the early 1960s vs Figure 5 (b) - a photograph of the area from the year 2017.

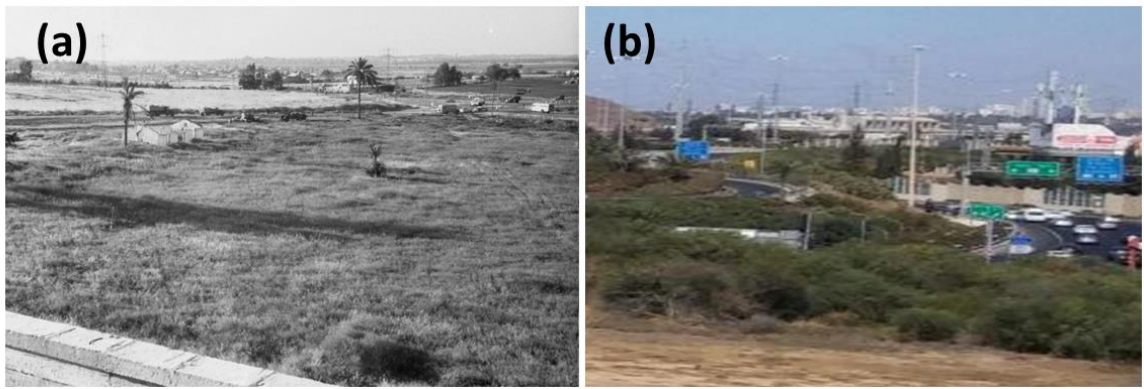

**Figure 5. The experiment area, early 60's (a) vs 2017 (b) (Credit: IMS).**

We note that the specific location where the measuring station is situated (a sand lot) has not changed meaningfully over the years, however the changes in area land surfaces are clearly apparent.

Importantly, vegetation gave way to concrete and asphalt roads, buildings, and other structures – all surfaces that absorb, rather than reflect, solar energy. As a result, surface temperatures, as well as overall ambient temperatures have risen. Thus, the urbanization process in the area, i.e. the UHI effect is also a probable cause for the temperature increase (Rotem-Mindali et al., 2015), and the meaningful reduction in the creation of fog, as a result.

When mentioning UHI, it is important to note that urbanization influences fog in different ways (Klemm and Lin, 2015). An increase in the UHI effect is associated with air temperature increases especially at night, thus resulting in a decreased tendency for formation of radiation fog, as we also observe in the current research (Figure 3a). Further, UHI is often associated with decreased water vapor content in the air, which results in the same effect. In rural, or agricultural areas, though, the issue is more complex – Increased temperatures may lead to a decrease in fog creation due to the reduction in RH. On the other hand, enhanced evapotranspiration caused by those higher temperatures may lead to an increase in fog. Regardless, the verification of the process based on meteorological data is extremely challenging, as the temperature changes involved are quite small, and the associated variations of RH unmeasurably so (within the 99% < RH < 100% range).

Moreover, prior research has shown that under the assumption of equilibrium conditions, both increase in air temperature, and decrease in concentration of aerosol particles lead to a reduction in development of fog and its intensity. Klemm and Lin (2015) have shown that in their case study, an increase of 0.1 percent in temperature had an equivalent effect to a decrease of 10% in aerosol concentration, where reductions of fog were concerned. If urbanization, as was observed in the experimental area (Fig. 5), is associated with increased air pollution, then fog formation can be enhanced. As comprehensive air pollution measurements for the entire experiment period were not available to us, as part of this research, this aspect was not investigated. A recent research has shown, though, that when urbanization and aerosol-pollution act together, the inhibiting effect of urbanization on fog dominants the much weaker aerosol-promoting effect (Yan et al., 2020).

We note that the literature data as reviewed (e.g. Klemm and Lin, 2015) indicates that, overall, as urbanization increases, decreases in fog occur more frequently than increases, as was also the case here.

**Summary**

In this paper 5 decades of fog data from Israel's central coastal plain was analyzed. The measurements indicate a decrease in the incidence of fog creation and a decrease in the frequency of radiation fog when compared to advection fog events. The decreasing fog trends detected here are in line with fog trends that have been widely observed across different parts of the world (Chen et al., (2006); Vautard et al., (2009); Van Schalkwyk (2012)). An in-depth investigation of the possible reasons for the decreasing fog trends in the experimental area is beyond the scope of this work, and is left for future research. However, we have indicated several factors that may have a role in creating the trends we report here.

Naturally, there are uncertainty factors in carrying out the observations. Thus, for example, we note that for the database we used, the measurements were stored every 3 hour interval. It is possible, then, that relatively shorter fog events might have occurred in between these sample times, and therefore were not tallied in this research. Additionally, it is important to note that the instruments measuring the different parameters were updated over time. It is possible that the location of the instruments was changed slightly, that different human observers carried out visibility observations, etc.

The results of this work can form the basis for future research that could be conducted on fog life cycles in the area, and indicate, for the first time, the trend of reductions of fog in this region.

Over the last decade numerous studies point to the potential that lies in the use of data from prevalent technologies and the 'Internet of Things' (IoT) to enhance the ability to measure various environmental phenomena (Overeem et al., 2013; Mass et al., 2014; Harel et al., 2015; Alpert et al., 2016; Price et al., 2018; David, 2019; Kumah et al., 2020), improve weather prediction capabilities (e.g. Kawamura et al., 2017) including fog in particular (e.g. David and Gao, 2016; 2018). However, the precise forecasting of this phenomenon remains an unsolved challenge (Koračin, 2017). Achieving better insight into the different mechanisms of fog formation, maintenance, and dissipation may lead to better forecasting capabilities, and, as a result, better capacity to contend with the dangers associated with this phenomenon.

*Acknowledgments*. The authors are grateful to the Israeli Meteorological Service (IMS) for providing the data required for this research. We extend our special thanks and appreciation to Nir Stav (General Director of the Israeli Meteorological Service), Amit Savir (Senior Deputy Director for Operational Meteorology) and Amos Porat (Manager of Climatic Services).